\documentclass[aps,prx,twocolumn,superscriptaddress,longbibliography]{revtex4-1}
\usepackage[T1]{fontenc}
\usepackage[utf8]{inputenc}
\usepackage{graphicx,amssymb,amsmath,xcolor}
\usepackage[unicode=true,
 bookmarks=false,
 breaklinks=false,pdfborder={0 0 0},pdfborderstyle={},backref=false,colorlinks=true]
 {hyperref}
 \hypersetup{
 colorlinks=true,
 linkcolor=blue, 
 citecolor=blue,  
 urlcolor=blue}

\makeatother

\begin{document}
\title{Magnon-Phonon Dynamics in Multidimensional Antiferromagnetic Oxides } 
\author{Yogendra Limbu}
\affiliation{Department of Physics and Astronomy, University of Iowa, Iowa City, Iowa 52242, USA}
\author{Michael E. Flatt\'e}
\affiliation{Department of Physics and Astronomy, University of Iowa, Iowa City, Iowa 52242, USA}
\affiliation{Department of Applied Physics and Science Education, Eindhoven University of Technology, Eindhoven, The Netherlands}
\author{Durga Paudyal}
\affiliation{Department of Physics and Astronomy, University of Iowa, Iowa City, Iowa 52242, USA}

\begin{abstract}

Antiferromagnetic oxides offer a compelling paradigm for magnonics. Combining \textit{ab initio} and spin Hamiltonian modeling, we map $\text{Cr}_2\text{O}_3$ from bulk to monolayer limits, verifying stability via phonon spectra. Contrary to previously predicted half-metallic ferromagnetic monolayer phases, we conclusively demonstrate that monolayer and bilayer configurations preserve an antiferromagnetic semiconductor ground state despite systematic gap narrowing. Intersecting magnon and phonon branches drive strong hybridization, establishing low-dimensional $\text{Cr}_2\text{O}_3$ as an ideal platform for quantum spintronics.

\end{abstract}

\maketitle

Quantized spin waves, commonly known as magnons, play a central role in 
determining the dynamical properties of magnetic materials and have emerged as 
key elements in the development of magnonics, spintronics, and quantum information 
science~\cite{vzutic2004spintronics,chumak2015magnon,tabuchi2015coherent,limbu2025magnetic}. 
While ferrimagnetic (FIM) insulators such as yttrium iron garnet (YIG)~\cite{larsen1975defects} 
and lithium aluminum ferrite (LAFO)~\cite{tong2026direct} are well known for 
their ultra-low magnetic damping~\cite{cherepanov1993saga,Serga_2010}, antiferromagnetic (AFM) insulators have emerged as the premier frontier for high-speed hardware. By eliminating parasitic stray fields and demonstrating ultra-fast, sub-terahertz to terahertz collective spin dynamics, AFM architectures allow for dense nano-scale integration and enhanced thermal stability~\cite{baltz2018antiferromagnetic,jungwirth2016antiferromagnetic}. 
Chromium oxide (Cr$_{2}$O$_{3}$) represents a prototypical AFM insulator; 
featuring a wide band gap of approximately 3.4~eV and a room-temperature N\'{e}el 
temperature of $T_{\text{N}} = 307$~K~\cite{mcguire1956antiferromagnetism}, it inherently 
hosts high-frequency magnon excitations~\cite{alikhanov1969neutron} capable of pure 
spin-current generation. However, the microscopic description of these excitations 
is governed by a delicate competition between direct- and super-exchange magnetic 
interactions. Because of the complex underlying AFM lattice, a substantial discrepancy 
persists between historical inelastic neutron scattering data and modern \textit{ab initio} 
predictions~\cite{samuelsen1970inelastic,su2025interplay,shi2009magnetism,fechner2018magnetophononics}. Resolving these exchange constants with quantitative rigor is crucial to accurately 
mapping and predicting out-of-equilibrium spin transport.

In corundum-structured transition metal (TM) oxides, electronic behavior evolves 
systematically with the TM $3d$ valence shell, transitioning from the strongly localized 
charge-transfer regime of $\alpha$-Fe$_{2}$O$_{3}$~\cite{catti1995theoretical,catti1996electronic} 
to the spin-paired metallic tendency of Ti$_{2}$O$_{3}$~\cite{guo2012electronic}. Situated 
at the critical boundary of this Mott-Hubbard/charge-transfer crossover, bulk 
Cr$_{2}$O$_{3}$ crystallizes in a trigonal $R\bar{3}c$ space group~\cite{catti1996electronic}, 
where octahedral Cr$^{3+}$ ions order into an alternating $\uparrow\downarrow\uparrow\downarrow$ 
spin sequence along the $\langle 111 \rangle$ direction~\cite{corliss1965magnetic}. 
Excitingly, moving beyond bulk crystals, our current findings establish that 
Cr$_{2}$O$_{3}$ can be structurally stabilized down to the monolayer (ML) limit, where it 
adopts a honeycomb-kagome geometry. While earlier modeling predicted that this 
two-dimensional configuration transitions into a half-metallic ferromagnet with a spin-filtering effect~\cite{hashmi2020ising}, a more rigorous treatment of strong electronic correlations is imperative to settle its true low-dimensional 
ground state.

In this letter, by combining advanced density functional theory (DFT + $U$) validated 
by hybrid functionals with an effective spin-Hamiltonian formalism, we map the 
evolution of electronic and coherent spin dynamics in low-dimensional Cr$_{2}$O$_{3}$. 
In contrast to the previously predicted half-metallic ferromagnetic (FM) state in ML phase, we 
conclusively demonstrate that both ML and bilayer (BL) configurations preserve an 
AFM semiconductor ground state despite a systematic narrowing 
of the fundamental band gap. Strikingly, linear spin-wave theory reveals a 
dimensionality-driven transition in the collective modes: the characteristic 
dual-degenerate chiral magnon branches of the bulk and BL limits transform into 
a singular, highly distinct acoustic magnon branch in the ML limit. Crucially, 
all computed collective modes operate deep within the terahertz regime. Further, 
the intersecting energy profiles of these magnon dispersions and the underlying 
lattice phonon spectra uncover a regime of pronounced magnon-phonon hybridization. 
These findings position low-dimensional Cr$_{2}$O$_{3}$ as an ideal, stray-field-free 
platform for coherent spin-lattice manipulation, high-frequency transport, and 
scalable quantum spintronic devices.

\begin{figure}
\centering
\includegraphics[width=0.48\textwidth]{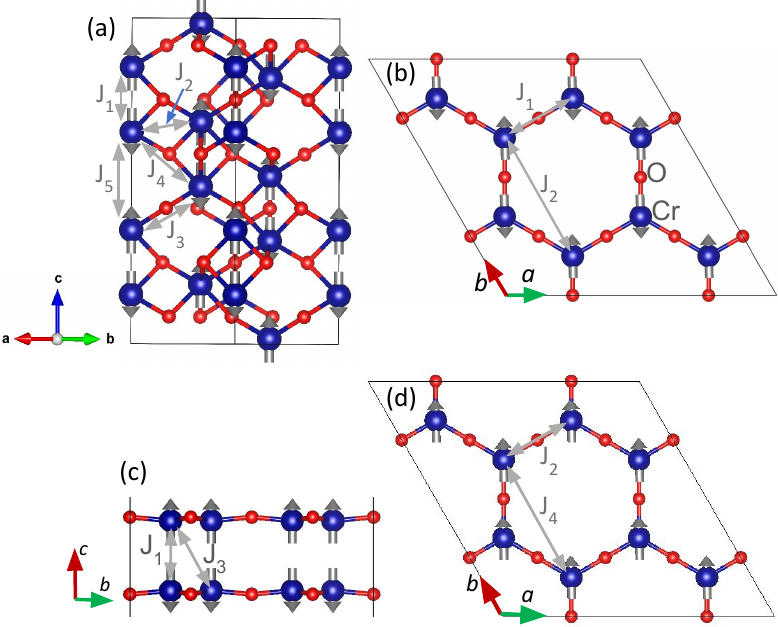}
\caption{Optimized crystal structures of Cr$_2$O$_3$. Bulk: (a) hexagonal conventional. $J_1$–$J_5$ denote the magnetic exchange interactions between first- to fifth-nearest-neighbor Cr atoms. ML: (b) $2\times2\times1$ supercell. BL: (c) and (d) are side and top views of AA stacking.}
\label{crystal_structure}
\end{figure}

We briefly outline the theory of spin-wave dynamics for calculating magnon dispersions. The spin Hamiltonian is written as,
$\mathcal{H}_\text{spin} = -\sum_{\langle i,j\rangle}J_{ij}\hat{S}_i . \hat{S}_j$,
where $J_{ij}$ denotes the magnetic exchange interaction between the $i^\text{th}$ and $j^\text{th}$ sites with spin operators $\hat{S}_i$ and $\hat{S}_j$. These exchange constants are used to compute the magnon dispersions by diagonalizing the spin Hamiltonian, followed by Holstein-Primakoff  transformation~\cite{holstein1940field,prabhakar2009spin}. This transformation allows to transfer the spin operators into bosonic operators. For complex magnetic materials, Colpa's formalism~\cite{colpa1978diagonalization} provides an efficient approach for diagonalizing the resulting Hamiltonian. The general quadratic bosonic Hamiltonian can be expressed as~\cite{de2023magnon}:

\begin{equation}
\mathcal{H} = \sum_k \psi_k^\dagger \mathcal{H}_k \psi_k
\label{para_H}
\end{equation}

\noindent where
$
\psi_{\textbf{k}}^\dagger =
\left(
\hat{a}^\dagger_{1,\textbf{k}},
\ldots,
\hat{a}^\dagger_{n,\textbf{k}},
\hat{a}_{1,-\textbf{k}},
\ldots,
\hat{a}_{n,-\textbf{k}}
\right)$, which satisfies the relation,
$\left[ \psi_{i, \mathbf{k}}, \psi_{i, \mathbf{k}'}^\dagger \right] =  \eta\, \delta_{\mathbf{k}, \mathbf{k}'}$

\noindent where 
\begin{equation}
\eta = \begin{pmatrix} 
\mathbb{I}_{n \times n} & \mathbb{O}_{n \times n} \\ 
\mathbb{O}_{n \times n} & -\mathbb{I}_{n \times n} 
\end{pmatrix} 
\nonumber
\end{equation}

\noindent Here, $\mathbb{I}_{n\times n}$ represents $n \times n$ identity matrix, whereas $\delta_{\textbf{k},\mathbf{k}'}$ is a Kronecker delta. The bosonic operators in momentum ($\mathbf{k}$) space are obtained from the real space bosonic operators through the Fourier transformation.

To obtain magnon dispersions, the Hamiltonian $\mathcal{H}_\mathbf{k}$ must be diagonalized by determining the transformation matrix $\mathcal{M}_\mathbf{k}$ that transforms the original bosonic basis $\psi_\mathbf{k}$ into a new basis $\phi_\mathbf{k}$, $\psi_\mathbf{k}$ = $\mathcal{M}_\mathbf{k}\phi_\mathbf{k}$,
where  $\phi_\mathbf{k}^\dagger = \begin{pmatrix} 
\alpha_{1, \mathbf{k}}^\dagger & 
\ldots, & 
\alpha_{n, \mathbf{k}}^\dagger &  
\alpha_{1, -\mathbf{k}} & 
\ldots, & 
\alpha_{n, -\mathbf{k}} 
\end{pmatrix}$. The new basis also satisfies the bosonic commutation relation.
Eq.~\ref{para_H} then becomes

\begin{equation}
\begin{aligned}
\mathcal{H} &= \sum_\mathbf{k} \phi_\mathbf{k}^\dagger \mathcal{M}_\mathbf{k}^\dagger \mathcal{H}_\mathbf{k} \mathcal{M}_\mathbf{k} \phi_\mathbf{k} \\
&= \sum_\mathbf{k} \phi_\mathbf{k}^\dagger \hat{\mathcal{H}}_\mathbf{k} \phi_\mathbf{k}
\end{aligned}
\label{dia_H}
\end{equation}

\noindent The matrix $\mathcal{M}_\mathbf{k}$ satisfies the relation, $\mathcal{M}_\mathbf{k} \eta \mathcal{\hat{M}}_\mathbf{k}^\dagger = \eta$. The eigenvalue of $\mathcal{H}_\mathbf{k}$ are obtained by diagonalizing $\eta\mathcal{H}_\mathbf{k}$.

First-principles calculations based on DFT are performed using the Vienna \textit{Ab initio} Simulation Package (\textsc{vasp})~\cite{furthmuller1996dimer, kresse1996efficiency}. The exchange-correlation interactions are treated within the generalized gradient approximation (GGA) parameterized by Perdew, Burke, and Ernzerhof (PBE)~\cite{perdew1996generalized}, alongside the projector-augmented wave (PAW) method~\cite{blochl1994projector}. The $\text{Cr}(3d^5 4s^1)$ and $\text{O}(2s^2 2p^4)$ configurations are explicitly treated as valence electrons, and the plane-wave kinetic energy cutoff is set to $500\text{~eV}$. Structural relaxations employ a $6 \times 6 \times 6$ $\Gamma$-centered $k$-point grid for bulk $\text{Cr}_2\text{O}_3$, and a $12 \times 12 \times 1$ grid for both the ML and BL phases. The total energy and ionic force convergence criteria are tightly set to $10^{-6}\text{~eV}$ and $10^{-5}\text{~eV}/\text{\AA}$, respectively. To eliminate spurious periodic interactions, a vacuum spacing of $20\text{~\AA}$ is enforced normal to the ML and BL planes. 

The electronic density of states (DOS) is evaluated via the Gaussian smearing method~\cite{jorgensen2021effectiveness}. To address the strong on-site Coulomb correlations among the localized Cr-$3d$ electrons and correct the characteristic GGA bandgap underestimation, we employ the $\text{GGA}+U$ approach with an effective Hubbard parameter $U_{\text{eff}} = 5\text{~eV}$. For a comprehensive phase analysis of the ML system, we additionally evaluate $U$ = 3~eV~\cite{limbu2025magnetic} and a formulation with $U = 2\text{~eV}$, $J = 0.7\text{~eV}$~\cite{hashmi2020ising}. Notably, standard GGA and GGA + $U$ ($U$ = 2~eV, $J$ = 0.7~eV) yield a half-metallic FM ground state, corroborating Ref.~\cite{hashmi2020ising}. Conversely, increasing the correlation to $U = 3$ or $5\text{~eV}$ stabilizes an AFM insulating ground state. This AFM phase is further benchmarked against HSE06 hybrid functional calculations~\cite{heyd2004efficient} incorporating 16\% exact Hartree-Fock exchange.

To accurately map the magnetic configurations, we construct a 2 $\times$ 2 $\times$ 1 supercell (48 Cr atoms) for the bulk, a 4 $\times$ 4 $\times$ 1 supercell (32 Cr atoms) for the ML, and a 2 $\times$ 2 $\times$ 1 supercell (16 Cr atoms) for the BL. To confirm the elimination of finite-size effects, test calculations for the BL using an expanded 3 $\times$ 3 $\times$ 1 supercell yield nearly identical exchange constants. Finally, magnetic exchange interactions are extracted via the four-state energy mapping technique~\cite{limbu2025stability, xiang2011predicting}, spanning up to the fifth-nearest-neighbors for the bulk and up to the fourth and second-nearest-neighbors for the BL and ML phases.

The optimized lattice parameters of bulk \(\text{Cr}_2\text{O}_3\) are \(a = b = 4.95\text{ \AA}\) and \(c = 13.81\text{ \AA}\), which agree well with available experimental and previous DFT results~\cite{catti1996electronic,greenwald1956changes,lebreau2014structural}. The calculated first- and second-nearest-neighbor \(\text{Cr--Cr}\) bond lengths are \(2.64\text{ \AA}\) and \(4.27\text{ \AA}\), showing excellent agreement with experiments~\cite{finger1980crystal}. Each \(\text{Cr}\) atom is coordinated by six \(\text{O}\) atoms to form \(\text{CrO}_{6}\) octahedra, featuring three shorter \(\text{Cr--O}\) bonds (\(1.97\text{ \AA}\)) and three longer \(\text{Cr--O}\) bonds (\(2.03\text{ \AA}\))~\cite{finger1980crystal,lebreau2014structural,rohrbach2004ab}. The primitive cell contains four \(\text{Cr}\) atoms with alternating spin-up and spin-down configurations along the $\langle 111 \rangle$ direction. The conventional cell displays the corresponding \(\text{Cr}\) spin configuration, with the \(\text{Cr}\) spins aligned antiparallel along the \(z\)-direction (Fig.~\ref{crystal_structure}a). 

\begin{figure}
\centering
\includegraphics[width=0.48\textwidth]{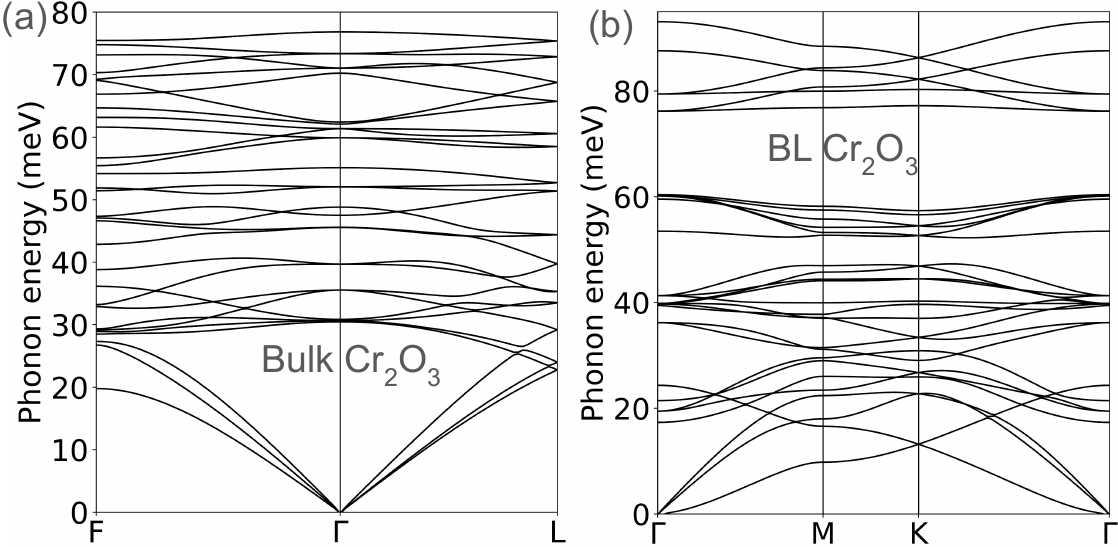}
\caption{Phonon dispersions of (a) bulk and (b) BL  (AA staking) Cr$_2$O$_3$. The positive phonon modes confirm the dynamical stability.}
\label{phonon}
\end{figure}

In contrast to bulk, the ML contains two Cr atoms in its primitive cell, with a Cr–Cr bond length of 3.63~\AA. Each Cr atom is coordinated by three equivalent O atoms with bond length of 1.81~\AA, forming two-dimensional honeycomb-kagome. Further, the BL structures are constructed by stacking two monolayers (MLs) with different relative stacking configurations. In the first configuration, corresponding to AA stacking, the two MLs are placed directly on top of each other, such that the atomic positions in the upper layer are vertically aligned with those in the lower layer. The AA-stacked BL has an optimized in-plane lattice constant of 6.20~\AA~ and an inter-layer separation of 3.03~\AA~ (Figs.~\ref{crystal_structure}c and ~\ref{crystal_structure}d). The material energetically favors inter-layer AFM coupling, with FM alignment of Cr within each layer. The intralayer Cr--Cr and Cr--O bond lengths are 3.58~\AA~ and 1.80~\AA, shorter than those in  ML. In the second configuration, the upper ML is laterally shifted relative to the lower ML such that the Cr atoms of one layer are positioned above the centers of the honeycomb hexagons of the other layer. This stacking configuration yields an inter-layer separation of 2.89~\AA, smaller than that of the AA-stacked BL. The optimized in-plane lattice constant is 6.24~\AA, while the intralayer Cr--Cr and Cr--O bond lengths are 3.60~\AA~ and 1.81~\AA. It is also favored inter-layer AFM, as AA-stacking.

The structural and chemical stabilities of these materials are assessed from their cohesive and formation energies~\cite{limbu2022electronic,limbu2025stability}, $E_{f/c} = E_\text{tot} - n_\text{Cr}E_\text{Cr}^\text{bulk} - n_\text{O}E_\text{O}^\text{bulk}$, where $E_\text{tot}$ is the total energy of the pristine phase. $n_\text{Cr}$ and $n_\text{O}$ represent the number of Cr and O atoms. $E_\text{Cr}^\text{bulk}$ and $E_\text{O}^\text{bulk}$ are the bulk energies of Cr and O atoms. The cohesive and formation energies of bulk and ML Cr$_2$O$_3$ are $-5.63$ and $-1.98$~eV/atom, and $-4.99$ and $-1.33$~eV/atom, respectively, confirming their phase stability. For AA-stacked and shifted BB-stacked BL Cr$_2$O$_3$, the cohesive and formation energies are $-0.545$ and $-4.15$~eV/atom, and $-0.45$ and $-4.06$~eV/atom, respectively, indicating structural and chemical stability in both configurations.

Further, the calculated phonon dispersions exhibit no imaginary modes in either the AFM or FM configuration (Fig.~\ref{phonon}a), verifying the dynamical stability of the bulk phase. The bulk phase exhibits 30 phonon modes, comprising 3 acoustic and 27 optical modes. The irreducible representations for the optical modes are expressed as, $\Gamma_{\mathrm{opt}} = 2A_{1g} + 5E_g + 3A_{2g} + 2A_{1u} + 2A_{2u} + 4E_u$ , where $E_u$ and $A_{2u}$ are IR active modes. $E_g$ and $A_{1g}$ modes are Raman active, whereas $A_{1u}$ and $A_{2g}$ are silent modes. The $E_g$ and $E_u$ modes are doubly degenerate. The largest phonon-frequency shift between the AFM and FM configurations is 8.92~meV for the $E_g$ mode, highlighting the significant spin-lattice coupling. This frequency shift is directly related to the spin-phonon coupling constant $\lambda$, defined by $\Delta\omega = \lambda\langle\mathbf{S}_i\cdot\mathbf{S}_j\rangle$, where $\langle\mathbf{S}_i\cdot\mathbf{S}_j\rangle$ represents spin-correlation function~\cite{limbu2025magnetic,olsson2021spin}. For Cr$^{3+}$, we take $S=3/2$, yielding $\lambda=3.97~\mathrm{cm}^{-1}$, significantly smaller than the calculated value for MXenes~\cite{limbu2025magnetic} but approximately twice the measured value for YIG~\cite{olsson2021spin}, potentially low magnetic damping favorable for energy-efficient spintronic applications. For ML Cr$_2$O$_3$, phonon frequencies are positive, confirming its dynamical stability~\cite{hashmi2020ising}. We similarly find that BL Cr$_2$O$_3$ with AA stacking is dynamically stable (Fig.~\ref{phonon}b), whereas the BB stacking is dynamically unstable despite its negative cohesive and formation energies. The BL unit cell contains four Cr and six O atoms, yielding 30 phonon modes. The irreducible representations for the optical modes are expressed as, $\Gamma_{\mathrm{opt}} =
2A_{1g}+2B_{1g}+B_{2g}+3E_{1g}+2E_{2g}
+A_{2u}+B_{1u}+2B_{2u}+2E_{1u}+2E_{2u}$. Here, $A_{1g}$, $E_{1g}$, and $E_{2g}$ are Raman-active modes, while $A_{2u}$ and $E_{1u}$ are IR-active modes. The remaining $B_{1g}$, $B_{2g}$, $B_{1u}$, $B_{2u}$, and $E_{2u}$ modes are silent. The $E_{1g}$, $E_{2g}$, $E_{1u}$, and $E_{2u}$ modes are doubly degenerate.

\begin{figure}
\centering
\includegraphics[width=0.42\textwidth]{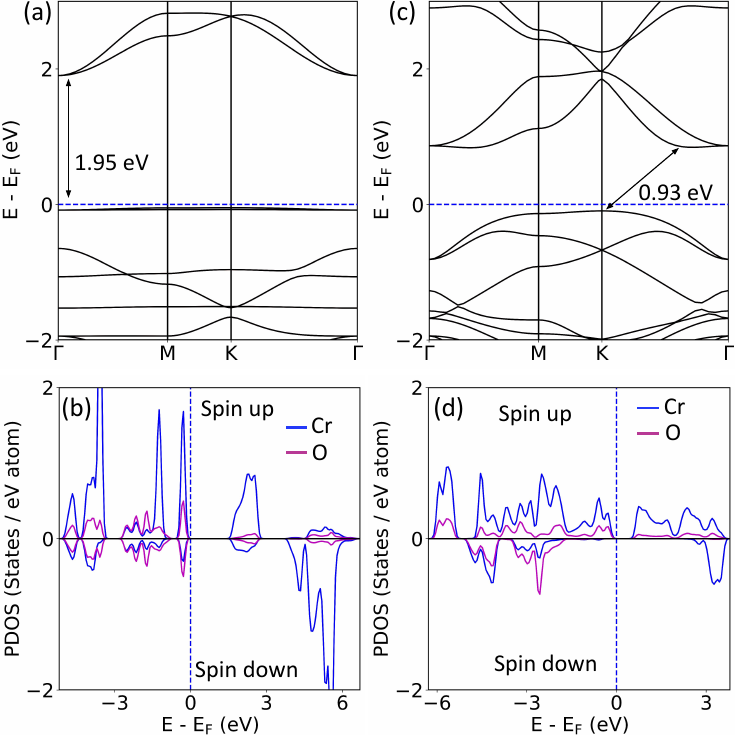}
\caption{Electronic band structures (a, c) and atom projected density of states  (b, d) of ML (a, b) and BL (c, d) Cr\({}_{2}\)O\({}_{3}\), calculated using GGA + \(U\) (\(U = 5\) eV).}
\label{electronic_structure}
\end{figure}

\begin{figure}
\centering
\includegraphics[width=0.49\textwidth]{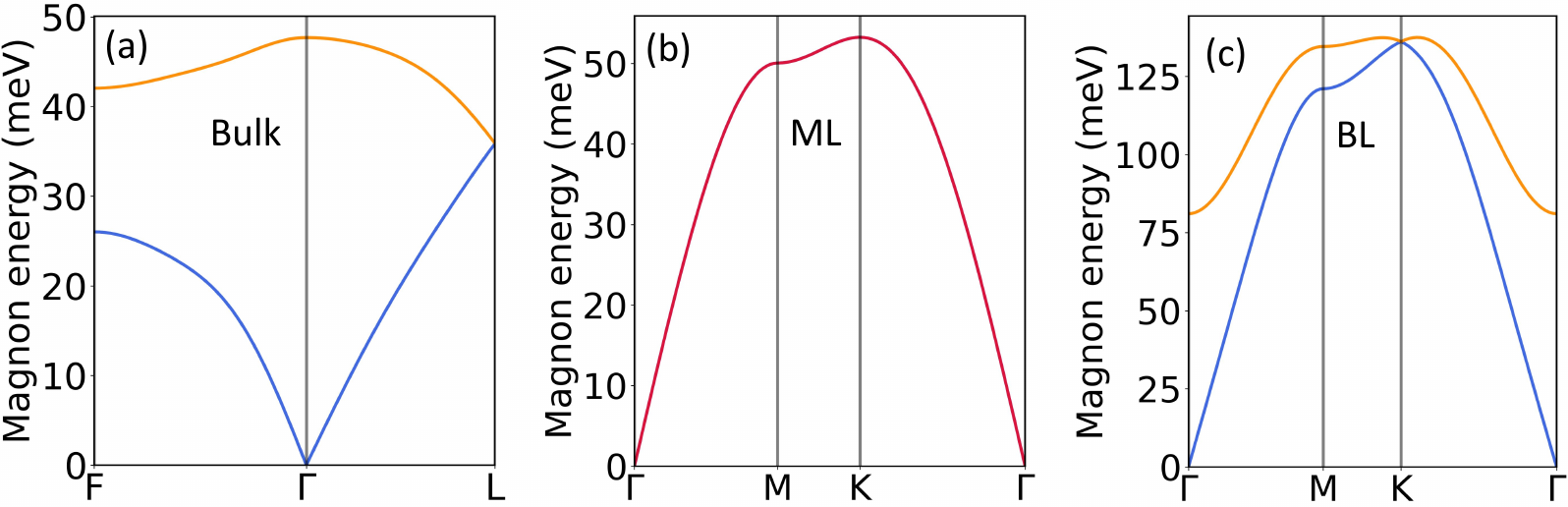}
\caption{Magnon dispersions of bulk (a), ML (b), and BL Cr$_2$O$_3$. Magnetic exchange interactions up to the fifth-, second-, and fourth-nearest-neighbor \(\text{Cr}\) atoms are included for the bulk, ML, and BL phases, respectively.}
\label{magnon}
\end{figure}

With GGA, bulk Cr$_2$O$_3$ is an AFM insulator with a band gap of 1.59~eV, substantially smaller than the experimental value of 3.4~eV~\cite{zimmermann1996electron}. Applying an effective Hubbard $U$ = 5~eV to the Cr-$3d$ electrons increases the band gap to 3.38~eV, consistent with the HSE result of 3.36~eV. Both values are in excellent agreement with experiment and previous DFT results~\cite{rohrbach2004ab} (Fig.~\ref{electronic_structure}a). Notably, the standard functional (HSE06 ), incorporating 25\% exact Hartree--Fock exchange and 75\% semilocal GGA exchange, overestimates the bulk Cr$_2$O$_3$ band gap, yielding 4.39~eV. Top of the valence band is primarily contributed from both Cr and O atoms, however, the bottom of the conduction band is mainly contributed from the Cr-$3d$ states, as seen in atom projected DOS. The computed spin magnetic moment (MM) for each Cr is 3.02~$\mu_B$. Using $U=5$~eV, the ML phase is an AFM insulator with a band gap of 1.95~eV (Fig.~\ref{electronic_structure}), slightly smaller than that of the bulk phase. This finding challenges the half-metallic FM ground state previously reported for this system~\cite{hashmi2020ising}. While that state can be reproduced using a weak Coulomb correlation ($U = 2$~eV, $J = 0.7$~eV), incorporating a more realistic Hubbard parameter of $U = 3$~eV~\cite{limbu2025magnetic} transitions the system to an AFM insulating ground state. Crucially, non-local HSE hybrid functional calculations consistently predict an AFM insulating ground state for both the bulk and ML phases. This independent verification demonstrates that a lower Hubbard value ($U = 2$~eV) fails to accurately capture the strongly localized Cr-$3d$ electrons, whereas the AFM state remains robust for $U = 3$ and $5$~eV within the GGA + $U$ framework. Quantitatively, the calculated total energy differences between the AFM and FM configurations ($\Delta E = E_{\mathrm{AFM}} - E_{\mathrm{FM}}$) are $0.28$~eV, $-0.15$~eV, and $-0.36$~eV per unit cell for GGA, GGA + $U$ ($U = 5$~eV), and HSE, respectively. The negative $\Delta E$ values unambiguously confirm the thermodynamic stability of the AFM phase, a result further underpinned by our magnetic exchange interaction analysis. Consequently, all subsequent calculations are executed within the validated GGA + $U$ ($U = 5$~eV) framework.

Interestingly, a relatively flat band appears just below the Fermi level, arising primarily from hybridized Cr-$3d$ and O-$2p$ states. As in the bulk phase, the valence-band maximum is dominated by Cr-$3d$ and O-$2p$ states, whereas the conduction-band minimum is primarily composed of Cr-$3d$ states. The calculated spin MM is 3.08~$\mu_B$/Cr. Furthermore, the BL retains its AFM ground state, with a reduced indirect band gap of 0.93~eV between the K and $\Gamma$ points (Fig.~\ref{electronic_structure}). A nearly flat valence band appears along the M--K path, as in ML Cr$_2$O$_3$. The electronic states near the Fermi level, including the valence-band maximum and conduction-band minimum, are predominantly derived from Cr states. The calculated spin MM of each Cr atom is 3.13~$\mu_B$, slightly larger than those of the bulk and ML phases.

To elucidate the microscopic origin of the magnetic order across different
dimensionalities, we map our total energy calculations onto a Heisenberg
Hamiltonian using an effective spin $S = 3/2$ for the $\text{Cr}^{3+}$ ions. We 
systematically evaluate the isotropic magnetic exchange interactions $J_i$ for 
bulk, ML, and AA-stacked BL $\text{Cr}_2\text{O}_3$. The 
calculated exchange constants are summarized in Table~\ref{bond_exchange}.

For the bulk phase, we consider interactions up to the fifth-nearest-neighbor ($J_1$--$J_5$). Our calculated exchange parameters exhibit excellent  quantitative agreement with inelastic neutron scattering data~\cite{samuelsen1970inelastic}. Specifically, the negative values of $J_1$ and $J_2$ signify robust  AFM coupling between the first- and second-nearest-neighbor $\text{Cr}$ pairs. Conversely, the positive values of $J_3$, $J_4$, and $J_5$  indicate FM alignment for the third- through fifth-nearest-neighbor pairs. While super-exchange mediated by the oxygen $p$-orbitals typically dictates the magnetic behavior in transition metal oxides, the magnetic ground state of $\text{Cr}_2\text{O}_3$ is governed by a subtle competition between direct and indirect exchange pathways. For the nearest-neighbor pairs, the face-sharing geometry of the oxygen octahedra reduces the $\text{Cr}$--$\text{Cr}$ distance sufficiently to induce a strong, direct $3d$--$3d$ orbital overlap. This direct exchange interaction overcomes the competing super-exchange pathways, ultimately driving the core collinear AFM spin arrangement~\cite{limbu2025magnetic, shi2009magnetism}.

Reducing the dimensionality to the ML limit drastically alters the 
structural symmetry and the local coordination environment, truncating the long-range exchange network. For ML $\text{Cr}_2\text{O}_3$, considering interactions up to the second-nearest-neighbor is sufficient to capture the  essential magnetic physics. We find that $J_1$ remains strongly negative, which is predominantly driven by AFM super-exchange with expected 180$^\circ$ angle, establishing a robust intralayer AFM ground state, whereas $J_2$  switches to or maintain FM coupling. Notably, the dominant magnitude of the intralayer $J_1$ stems from a contractive relaxation of the first-nearest-neighbor $\text{Cr}$--$\text{Cr}$ distance in the 2D limit. This enhanced exchange coupling pushes the magnon excitations into the few-terahertz regime, highlighting the potential of ML $\text{Cr}_2\text{O}_3$ for high-speed AFM spintronics.

We now explore the effect of vertical confinement and inter-layer hybridization in AA-stacked BL $\text{Cr}_2\text{O}_3$ by expanding the exchange network up to the fourth-nearest-neighbor. The out-of-plane stacking introduces strong inter-layer channels without significantly disturbing the underlying intralayer configurations. The primary inter-layer coupling, quantified by a 
remarkably large negative $J_1$, reveals a strong AFM interaction between the vertically adjacent $\text{Cr}$ sites. Inter-layer AFM coupling is further reinforced by a negative $J_3$ between third-nearest-neighbor pairs. In contrast, the in-plane magnetic integrity is preserved via the positive $J_2$ and $J_4$ constants, which enforce FM intralayer coupling between the second- and fourth-nearest-neighbor sites. This delicate hierarchy of strong AFM inter-layer and competing FM/AFM intralayer couplings underscores the highly tunable nature of the magnetic order in $\text{Cr}_2\text{O}_3$ van der Waals materials.

\begin{table}[t]
\caption{Magnetic exchange
interactions for bulk, ML, and BL
Cr$_2$O$_3$ in meV. The computed exchange constants  are compared with available experimental results.}
\label{tab:exchange}
\begin{ruledtabular}
\begin{tabular}{lccccc}
System & $J_1$ & $J_2$ & $J_3$ & $J_4$ & $J_5$ \\
\hline
Bulk & -6.83 & -4.78 & 1.67 & 1.57  & 0.15 \\
Exp.~\cite{samuelsen1970inelastic} & -7.52 & -3.26 & 0.06 &  0.017 & -0.19 \\
ML & -9.12 & 0.90 & -- & --   & --   \\
BL & -37.10 & 3.46 & -1.67 & 5.06 & --\\
\end{tabular}
\label{bond_exchange}
\end{ruledtabular}
\end{table}

Using the extracted exchange constants, we compute the spin-wave spectra to elucidate the collective magnetic excitations across different dimensionalities. For the bulk phase, the magnon dispersion hosts two distinct branches, reaching a maximum excitation energy of 47.70~meV at the $\Gamma$ point. These modes are in excellent agreement with both previous time-dependent density functional theory (TD-DFT) calculations~\cite{skovhus2022magnons} and inelastic neutron scattering measurements~\cite{samuelsen1970inelastic}. Notably, the acoustic magnon exhibits an isotropic-to-anisotropic crossover: it disperses approximately linearly along the $\Gamma$--L direction but displays a nearly quadratic ($\propto k^2$) dependence along the $\Gamma$--F path, reflecting the directional dependence of the underlying exchange anisotropy. 

Reducing the dimensionality to a ML drastically alters the excitation spectrum. The mapped exchange interactions yield a single magnon branch, a direct consequence of the compensated AFM ordering between the two oppositely aligned Cr atoms within the primitive cell. The ML spectrum reaches a peak energy of 53.24~meV at the high-symmetry K point. In stark contrast to the bulk phase, the ML dispersion remains nearly linear along both the $\Gamma$--K and $\Gamma$--M paths, which is a classic hallmark of a two-dimensional isotropic Heisenberg antiferromagnet.

The BL $\text{Cr}_2\text{O}_3$ introduces compelling cross-layer physics and topological features. We obtain two magnon branches featuring an exotic topological degeneracy at the K point at an energy of 136.14~meV. While the intrinsic Dzyaloshinskii-Moriya interaction (DMI) is expected to break inversion symmetry and open a topological gap at this vertex, it is neglected here to isolate the pure exchange-driven band structure. Remarkably, the highest magnon excitation shifts to the K-$\Gamma$ direction, peaking at an elevated energy of 137.47~meV. This pronounced energy enhancement relative to both the bulk and ML counterparts is driven by intense inter-layer coupling; specifically, it stems from strong magnetic exchange between first-nearest-neighbor Cr atoms mediated via direct vertical Cr--Cr orbital overlap. This robust inter-layer coupling shifts the optical spectral weight toward higher frequencies, positioning BL $\text{Cr}_2\text{O}_3$ as a compelling platform for ultra-fast spin dynamics and high-frequency magnonic devices.
Meanwhile, the lower-energy acoustic branch retains a linear dispersion along $\Gamma$--K, confirming that robust intra-layer AFM correlations remain intact despite strong inter-layer hybridization.

In conclusion, our systematic \(\textit{ab\ initio}\) calculations reveal AFM oxides to be a versatile material class for energy-efficient magnonics. By integrating advanced DFT with effective spin Hamiltonian modeling, we have mapped the evolution of the electronic structure, exchange interactions, and collective excitations of Cr$_2$O$_3$ across the bulk, BL, and ML limits. Our thermodynamic and lattice dynamics calculations confirm the robust structural stability of these multidimensional configurations, supported by favorable cohesive energies and positive phonon frequencies. Benchmarking the localized Cr-$3d$ Hubbard $U$ correction against hybrid functionals provides a rigorous framework for accurately capturing electronic properties in correlated oxides. Contrary to prior predictions of a half-metallic FM state in ML phase, we demonstrate that despite a systematic band-gap narrowing caused by dimensional reduction, both ML and BL systems retain a stable AFM semiconductor ground state. Beyond electronic structure, our linear spin-wave analysis uncovers a striking dimensionality-dependent tuning of the collective modes. We resolve a clear transition from the two-fold magnon branches characteristic of the bulk and BL geometries to a single, distinct magnon branch in the ML limit, with all excitations operational within the high-speed terahertz regime. Crucially, the spatial overlap between these magnon dispersions and the underlying phonon spectra points to a prominent magnon-phonon coupling. This intrinsic hybridization not only offers a potential for coherent spin-lattice manipulation but also suggests new avenues for exploring non-equilibrium magnon transport and thermal spin currents in atomically thin devices. Ultimately, these results firmly establish multidimensional Cr$_2$O$_3$ as a premier material for next-generation, high-frequency spintronics, serving as a critical blueprint for the design of scalable, stray-field-free spintronic and quantum information technologies.

This work was supported as part of the Center for Energy Efficient Magnonics, an Energy Frontier Research Center funded by the U.S.\@ Department of Energy, Office of Science, Basic Energy Sciences, under Award number DE-AC02-76SF00515. D.P acknowledges the use of the computational facilities on the Frontera supercomputer at the Texas Advanced Computing Center (TACC) via the pathway allocation, DMR23051.

\bibliography{references}
\end{document}